\documentclass[twocolumn,pre,aps,superscriptaddress,10pt]{revtex4-1}

\usepackage{amsfonts,amssymb}
\usepackage[sumlimits,intlimits]{amsmath}
\usepackage{graphics}
\usepackage{graphicx}
\usepackage{mathrsfs}
\usepackage{verbatim}
\usepackage{hyperref}    
\usepackage{bm}          

\pdfoutput=1

\begin{document}

\title{A model of large volumetric capacitance in graphene supercapacitors based on ion clustering}

\date{\today}

\author{Brian Skinner}
\affiliation{Fine Theoretical Physics Institute, University of Minnesota, Minneapolis, MN 55455, USA}
\author{M. M. Fogler}
\affiliation{Department of Physics, University of California-–San Diego, 9500 Gilman Drive, La Jolla, CA 92093, USA}
\author{B. I. Shklovskii}
\affiliation{Fine Theoretical Physics Institute, University of Minnesota, Minneapolis, MN 55455, USA}

\begin{abstract}

Electric double layer supercapacitors are promising devices for high-power energy storage based on the reversible absorption of ions into porous, conducting electrodes.  Graphene is a particularly good candidate for the electrode material in supercapacitors due to its high conductivity and large surface area.  
In this paper we consider supercapacitor electrodes made from a stack of graphene sheets with randomly-inserted ``spacer'' molecules.  We show that the large volumetric capacitances $C \gtrsim 100$ F/cm$^3$ observed experimentally can be understood as a result of collective intercalation of ions into the graphene stack and the accompanying nonlinear screening by graphene electrons that renormalizes the charge of the ion clusters.

\end{abstract}
\maketitle

\section{Introduction} \label{sec:intro}

Electric double layer supercapacitors (SCs) are a promising class of efficient, long-lasting, high-power electrical energy storage devices based on reversible adsorption of ions onto the surface of a porous electrode.  The most common SC devices use activated carbons as electrode materials because of their high specific surface area and moderate cost.
Recent improvements in electrode materials have greatly improved the energy storage capacity of SCs, so that SCs are fast becoming viable candidates to replace conventional batteries in a number of energy storage applications \cite{Abruna2008bae, Schindall2007cou}.

\begin{figure}[htb!]
\centering
\includegraphics[width=0.5 \textwidth]{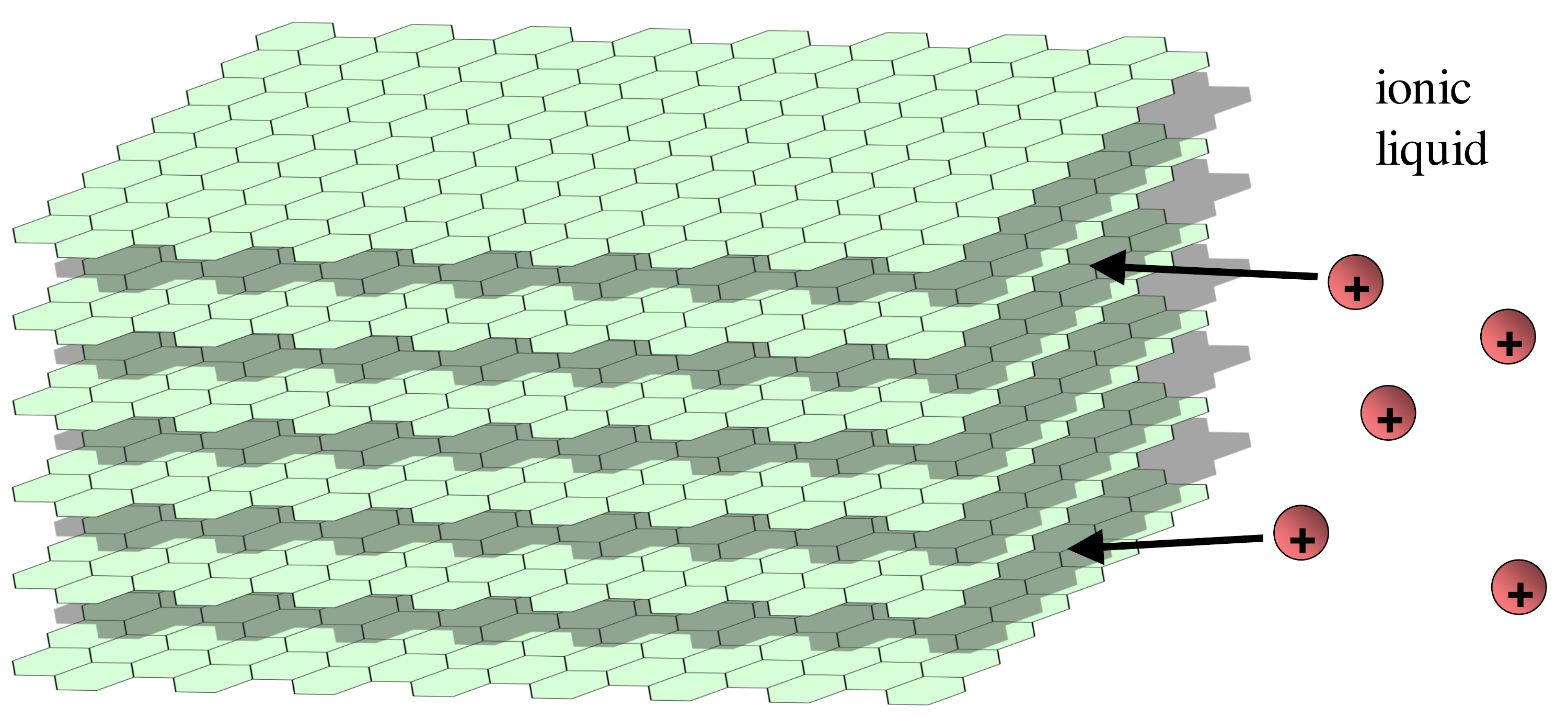}
\caption{(Color online) A schematic depiction of a graphene supercapacitor electrode.  Graphene sheets are arranged in a stack that is placed in contact with a reservoir of ionic liquid.  As a voltage is applied between this electrode (which we take to be the negative electrode) and the opposite electrode (not shown), cations (red circles with +'s) are driven to intercalate between graphene layers to neutralize the electrode's negative electronic charge.} \label{fig:schematic}
\end{figure}

In this paper we study SC devices made from graphene electrodes, which have attracted considerable attention in recent years \cite{Zhang2010gma, Zhang2010gnc, Kim2011hsb, Stoller2008gu, Wang2009sdb, Vivekchand2008ges, Simon2008mec}.  In such devices individual graphene layers are stacked to form an electrode and placed in contact with a reservoir of ionic liquid (or some other concentrated electrolyte solution), as shown schematically in Fig.\ \ref{fig:schematic}.  A functional SC device contains two such electrodes with a voltage applied between them, usually with a porous separator between them that isolates the electrodes from each other electronically while allowing ions to flow from the reservoir to either electrode.  

Experiments on graphene-based SCs have demonstrated~\cite{Stoller2008gu, Vivekchand2008ges, Zhang2010gma} volumetric capacitances in excess of $100\,\text{F}/\text{cm}^3$. These large capacitances are usually explained within the simple picture in which the capacitor energy is stored in the electric double layer (EDL) that forms when ions are adsorbed onto the electrode's active surface. The volumetric capacitance of the electrode is then written as
\begin{equation}
C = A \mathscr{C}_{\text{EDL}}\,,
\label{eq:C_EDL}
\end{equation}
where $\mathscr{C}_{\text{EDL}}$ is the EDL capacitance per unit area and $A$ is the surface area per unit volume of the electrode. 
In order to make a conservative estimate of the value of $\mathscr{C}_{\text{EDL}}$ necessary to explain the large experimental values of capacitance, one can imagine that an EDL forms on both sides of every graphene sheet and that adjacent graphene sheets have just enough separation to allow ions to fit sterically between them.  In this case for an ion diameter $a = 1$ nm one finds that the value $100 $ F/cm$^3$ implies an EDL capacitance of at least $\mathscr{C}_{\text{EDL}} \approx 7$ $\mu$F/cm$^2$.  This value of $\mathscr{C}_\text{EDL}$ is similar to the reported EDL capacitances of a metal/ionic liquid interface \cite{Baldelli2008ssa, Islam2008eds}.  Thus, to explain the large experimental value of $C$ one should apparently imagine that each graphene sheet provides two independent EDLs whose capacitance is as large as that of a free metal/ionic liquid interface. 

Such large capacitance of graphene SCs is difficult to understand if one recalls that graphene is not a metal but a semi-metal. Graphene has a relatively small thermodynamic density of states (TDOS) $\nu = {d n_e} / {d \mu_e}$, where
$n_e$ is the two-dimensional (2D) concentration of electrons and 
$\mu_e$ is their chemical potential, which is tuned by the capacitor voltage.  It is well known~\cite{Metrot1985rbc, Gerischer1987dot, Dahn1992dos, Xia2009moq, Ponomarenko2010dos} that within the mean-field theory this finite TDOS modifies the EDL differential capacitance as follows:
\begin{equation}
\mathscr{C}_{\text{EDL}}^{-1} = \mathscr{C}_{\text{EDL}, \infty}^{-1}
 + \mathscr{C}_q^{-1}\,,
\quad
\mathscr{C}_q \equiv e^2 {\nu}\,.
\label{eq:C_EDL_MF}
\end{equation}
That is, one effectively has a ``quantum capacitance'' $\mathscr{C}_q$ that adds in series with the ideal EDL capacitance $\mathscr{C}_{\text{EDL}, \infty}$ which is achievable only at the surface of a hypothetical perfect metal with infinite TDOS. Hence, the quantum capacitance imposes an upper bound on the volumetric capacitance of the system.

This statement has profound implications for graphene where the TDOS has the form
\begin{equation}
     \nu(\mu_e, T) = \frac{4}{\pi}\, \frac{k_BT}{\hbar^2 v^2}\, \ln \left(
             2 \cosh \frac{\mu_e}{2 k_BT} \right)\,,
\label{eq:nu_T}
\end{equation}
where $k_BT$ is the thermal energy and $v = 1.0 \times 10^6\,\text{m} / \text{s}$ is the Fermi velocity. The corresponding quantum capacitance has a deep minimum at the neutrality point $\mu_e = 0$ equal to
\begin{equation}
\mathscr{C}_{q, \text{min}} = \frac{4 \ln 2}{\pi}\, \frac{e^2}{\hbar^2 v^2}\, k_BT = 0.83\, \mu\text{F} / \text{cm}^2
\label{eq:C_q_min}
\end{equation}
at $T = 295\,\text{K}$. Therefore, the mean-field theory of Eq.\ \eqref{eq:C_q_min} predicts i) a minimum value of $C$ that is a factor of eight smaller than what is observed and ii) a strongly varying $U$-shaped voltage dependence of the capacitance, which is at odds with the roughly constant measured $C(V)$.

These same inconsistencies have been discussed in Ref.~\onlinecite{Xia2009moq}, where an EDL capacitance $\mathscr{C}_{\text{EDL}} = 7$--$10\,\mu\text{F} / \text{cm}^2$ was measured at the interface of a single graphene sheet and ionic liquid. It was suggested that graphene in this experiment was subject to randomly positioned charged impurities, causing smearing of the ideal TDOS and an increased $\mathscr{C}_{q, \text{min}}$. By analogy, one can argue that multi-layered graphene SC devices~\cite{Zhang2010gma, Zhang2010gnc, Kim2011hsb, Stoller2008gu, Wang2009sdb, Vivekchand2008ges, Simon2008mec}, which are doubtless heavily disordered,
also have higher $\nu$ than Eq.~\eqref{eq:nu_T} predicts~\footnote{In principle, higher quantum capacitance can also result from coupling $\gamma_1$ between graphene layers. In bulk graphite, where $\gamma_1 = 0.4\,\text{eV}$ the volumetric
quantum capacitance is $C_{q, \text{min}} = (8 / \pi) \varepsilon_0 (\gamma_1 / e^2 c_0)(e^2 / \hbar v)^2 \approx 100\,\text{F}/\text{cm}^3$, in agreement with
experiments of lithium intercalation~\cite{Dahn1992dos}.}.

In this paper we explore a more intriguing possibility, namely that large enhancement of capacitance can result from a breakdown of the mean-field theory. We propose a theoretical model in which ions enter the electrode cooperatively as dense clusters, in analogy to staging in intercalated graphite compounds~\cite{Fischer1978gic, Dresselhaus1981ico}. The effective attraction between the like-charged ions is mediated by elastic stresses induced in the graphene stack~\cite{Safran1979lre, Safran1980csa, Ohnishi1980tei, McKinnon1980sei, Park1996eib}. The high charge concentration inside such ion clusters activates a strong nonlinear screening in layered graphene~\cite{Pietronero1978cdi, Safran1980eia, Safran1980pds} that renormalizes the ion charge. We show that as a result the volumetric capacitance is greatly enhanced above the mean-field value.

Our general approach to calculating the capacitance of a given electrode (say, the negative electrode) is as follows.
We first compute the total free energy $F(N_+)$ per unit volume associated
with the lowest energy configuration of $N_+$ cations (and the neutralizing concentration of $N_{+}$ electrons) per unit volume
in the electrode.  The value of
the electronic charge $Q = eN_+$ per unit volume of the electrode is that which minimizes the system's
total free energy density $F - QV$, where the term $-QV$ represents the work
done by the voltage source.
Using the equilibrium condition $d(F - QV)/dQ = 0$ gives 
\begin{equation}
V = \frac{d F}{d Q} = \frac{1}{e} \frac{d F}{d N_+}. 
\label{eq:Vdef} 
\end{equation} 
The differential capacitance per unit volume of the
electrode, $C = (dV/dQ)^{-1}$, can therefore be written 
\begin{equation} 
C = \left( \frac{d^2 F}{d Q^2} \right)^{-1} = e^2 \left( \frac{d^2 F}{d N_+^2} \right)^{-1}. 
\label{eq:Cdef} 
\end{equation} 
In this way, the capacitance is closely related to the charge compressibility $(N_+^2 d^2F/dN_+^2)^{-1}$ of the system: large capacitance implies high compressibility.  The capacitance can be expressed as a function of voltage, $C(V)$, by
combining Eqs.\ (\ref{eq:Vdef}) and (\ref{eq:Cdef}).
These relations give the capacitance of a single electrode (here, the negative electrode); the total capacitance of the SC device is the series sum of the anode and cathode capacitances.  In this paper we concern ourselves with the capacitance of a single electrode only.  For the majority of this paper we neglect entropic effects, so that in Eqs.\ \eqref{eq:Vdef} and \eqref{eq:Cdef} the free energy $F$ can be replaced by the total energy $U$.

The remainder of this paper is organized as follows. Section~\ref{sec:GSelectrodes} defines the model system to be studied. In Sec.~\ref{sec:CGS-MF} we discuss the capacitance of the graphene electrode in the mean-field approximation.  Section~\ref{sec:staging} overviews the physics of staging and nonlinear screening in graphite intercalation compounds and calculates their capacitance.
In Sec.~\ref{sec:alpha1} the mean-field and staging results for capacitance are united and the main results of this paper are derived.  We show that the capacitance of the SC everywhere exceeds the mean-field quantum capacitance.  Finally, in Sec.\ \ref{sec:GSdiscussion} we provide additional discussion of the large-voltage regime, at which the mean-field quantum capacitance manifests itself in a somewhat enhanced form, and we provide some concluding remarks.

\section{Graphene-based supercapacitor electrodes} \label{sec:GSelectrodes}

Over the past few years a number of studies have confirmed the large capacitance per unit volume of SC devices in which the electrode is made from a stack of graphene sheets \cite{Simon2008mec, Zhang2010gma, Zhang2010gnc, Kim2011hsb, Stoller2008gu, Wang2009sdb, Vivekchand2008ges, Zhu2011csp}, as depicted in Fig.~\ref{fig:schematic}.  Most commonly, in such devices the graphene sheets are ordered in the transverse direction, but have random translations or rotations between adjacent sheets (``turbostratic disorder''), so that on average the electronic dispersion relation of each graphene sheet is unaltered by inter-layer coupling.  Below we refer to this arrangement as a ``graphene stack'' (GS). In practice, the layered ordering of graphene sheets within the GS electrode is preserved only over some mesoscopic length scale, usually on the order of tens of nanometers \cite{Simon2008mec, Zhu2011csp}.   In the remainder of this paper we make the assumption that ordering of graphene planes within the electrode is preserved over arbitrarily large distances.  This assumption does not significantly affect any results, as we demonstrate in Sec.~\ref{sec:alpha1}.

%
%
\begin{figure*}[htb!]
\centering
\includegraphics[width=0.7\textwidth]{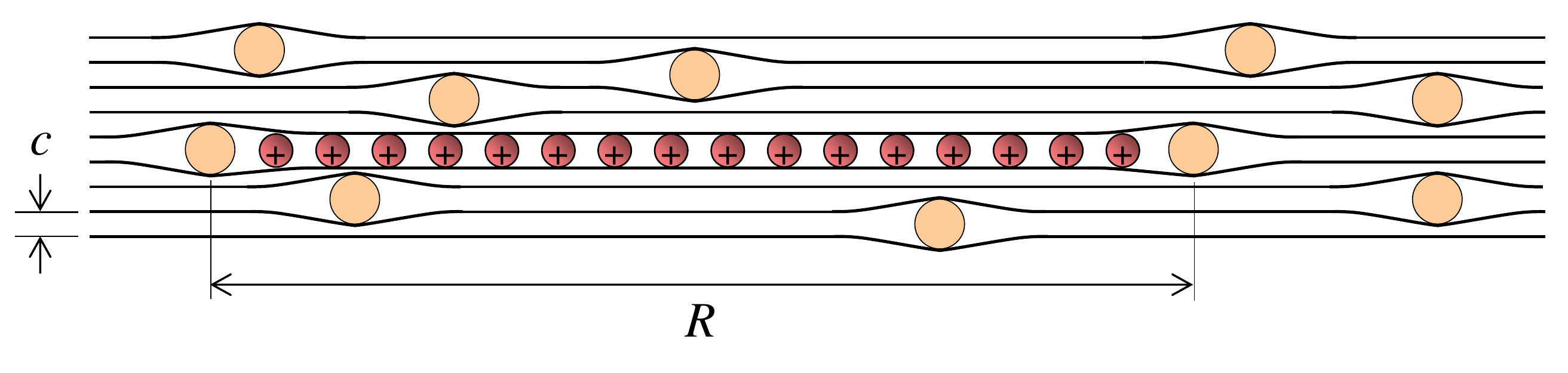}
\caption[Schematic picture of a graphene stack]{A schematic picture of a portion of the GS electrode, shown in a side view.  The stack of graphene sheets (black lines), arranged with a separation $c$ between layers, has some concentration of ``spacer'' molecules (tan-colored circles) in it.  When a voltage is applied, ions (red circles with $+$'s) enter the electrode collectively in the form of disks whose size is determined by the average distance $R$ between spacers in a given plane. } \label{fig:GS-schematic}
\end{figure*}

When the positive and negative electrodes are connected to opposite terminals of a voltage source, the applied voltage $V_{tot}$ between them is related to the system's free energy $F_{tot}$ according to $V_{tot} = dF_{tot}/dQ$.  In this paper, for simplicity, we focus only on the negative electrode, employing the definition of voltage given in Eq.\ \eqref{eq:Vdef}.  This voltage $V$ is related to the total voltage applied between the electrodes according to $V_{tot} = V + dF'/dQ$, where $F'$ is the free energy density of the configuration of $Q/e$ anions and electron holes within the positive electrode.  In this way the SC is equivalent to two capacitors connected in series, with the applied voltage $V_{tot}$ being shared between the two halves of the system.  One can also think that the results we derive below correspond to a device with a very large positive electrode, so that the smaller negative electrode's capacitance dominates the series sum and $V_{tot} \simeq V$.

When the voltage $V$ is applied, cations are driven to intercalate into the negative electrode and they establish a uniform electrochemical potential throughout the device.
In our treatment below we make the simplifying assumption that the central ionic liquid reservoir has a large negative chemical potential due to the strong, Coulomb-based correlation energy of ions within the ionic liquid reservoir, so that at $V = 0$ both electrodes are empty of ions.  In this case the capacitor acquires a finite charge only when $V$ is larger than some positive ``threshold voltage'' $V_t$.  This behavior has been described previously for dual-graphite energy cells \cite{Seel2000eio}.

For practical SC devices, the effectiveness of the electrode is often limited by the strong binding of graphene sheets to each other through van der Waals (vdW) attraction \cite{Zhang2010gma, Kim2011hsb}.  The short-ranged vdW interaction between sheets produces a large contact energy $\gamma = (1.5 \pm 0.4)$ eV/nm$^2$ \cite{Spanu2009nas}, and in the absence of any modification to the graphene this interaction results in dense, graphite-like agglomerates whose interior surface area is not readily accessible to ions.  In this case only a large applied voltage can induce ions to enter the electrode by forcibly separating the space between adjacent graphene sheets.  This process is often too slow to meet the fast charging/discharging requirements of practical devices.  Further, such agglomeration can require the device to operate at large voltages that are outside the limited electrochemical stability window of the electrolyte.  

One method of overcoming these difficulties is to fabricate electrodes from chemically modified graphene, in which graphene sheets are bonded to polymers or some other ``spacer'' molecule before being condensed into a stack similar to the one shown in Fig.~\ref{fig:GS-schematic}.  In this arrangement each spacer provides sufficient steric hindrance to separate adjacent graphene layers at a given point.  Electrodes with such spacers have recently been constructed and shown to generate capacitance values in excess of 100 F/cm$^3$ even at relatively low voltages \cite{Kim2011hsb}.  

In the region surrounding each spacer, separated graphene sheets experience a strong elastic stress that results from the competition between their attractive vdW interaction and the elastic energy associated with bending them to bring them together.  The introduction of intercalated ions into the GS requires additional separation of graphene sheets, since the diameter $a$ of the ions is also larger than the spacing $c$ between adjacent graphene sheets (and is typically comparable to the size of the spacers).  In equilibrium, this separation occurs in a way that best relieves the elastic stress surrounding the spacers --- namely, through the formation of filled, 2D ``pockets'' of ions rather than through the introduction of scattered, individual ions.  The size of these pockets is defined by the average distance $R$ between spacers (see Fig.~\ref{fig:GS-schematic}).

One can explain the formation of these pockets of ions using the language of elastic energy-mediated attraction.  When a single ion enters the GS, it creates a region of intense elastic stress around itself.  If two such ions are introduced to the GS, they can reduce the total elastic energy of the system by approaching each other closely, so that together they share the elastic energy associated with deforming the GS around them.  Thus, there is an effective attraction between ions in the GS mediated by the elastic strain they induce.  This attraction has been observed experimentally in intercalation experiments \cite{Dresselhaus1981ico, Ishihara2011iop} and has been described theoretically as an effective lateral attractive interaction \cite{Safran1980pds, Safran1980eia}.  

When there is some finite concentration of ions within the electrode, scattered ions between two adjacent sheets in the GS attract each other to form a large ``disk'' of ionic charge.  The growth of the disk is truncated at the size $R$, at which point ions fill the area between neighboring spacers.  These spacers are generally  massive enough that they can be considered immobile on the time scales of charging/discharging of the electrode.  The 2D charge density $\sigma_0$ of the disk is determined by the balance between the elastic energy-mediated attraction and the Coulomb repulsion between ions.

The result is that, as a voltage is applied, the electrode is charged by the incremental addition of disks of ions (which may have a somewhat irregular shape) that enter the electrode collectively, as shown in Fig.~\ref{fig:GS-schematic}.  In our derivation of the capacitance in Sec.\ \ref{sec:alpha1} we assume that the disk size $R \gg a > c$, so that every disk contains a large number of ions.

One may well notice that in the limit where there are no spacers at all in the GS, the electrode is simply turbostratic graphite.  In this case the disk size $R \rightarrow \infty$, which suggests that ions enter the GS as infinite, uniform planes.  In fact, this phenomenon is well-known: it is referred to as ``staging'' of graphite intercalation compounds  and has been studied for more than eighty years \cite{Fischer1978gic, Dresselhaus1981ico}.  In such compounds, guest ions such as Li$^+$, K$^+$, or Ca$^{2+}$ are forcibly introduced into pure graphite by an applied voltage.  In most cases, the elastic energy-mediated attraction causes the intercalated ions to arrange themselves in an ordered sequence of filled and empty interlayer galleries, with each filled ion layer having some fixed 2D concentration of ions and the distance between filled layers decreasing with the overall concentration of intercalated ions.  The number of graphene sheets between two subsequent filled layers is called the ``stage'' of the compound: low densities of intercalated ions corresponds to large stage, while the maximum filling of ions is called ``stage 1.''   The concentration $\sigma_0/e$ of ions within a filled plane depends in general on the ion size: small ions tend to have stoichiometry XC$_6$ in their stage 1 form, while larger ions form less dense arrangements such as XC$_8$ or XC$_{12}$ \cite{Huggins2009abm}.  In each case, $\sigma_0 /(e/c^2)$ is of order unity.  The physics of staging and its implications for capacitance are discussed more fully in Sec.~\ref{sec:staging}.

\section{Mean-field theory of capacitance in a graphene stack} \label{sec:CGS-MF}

As a first attempt at describing the capacitance of a GS, one can try to use mean-field theory.  This approach ignores any correlations among ions, such as the clustering described in Sec.~\ref{sec:GSelectrodes}.  In the simplest mean-field model, discrete ions are replaced by a uniform charge density $eN_+$ that fills the electrode volume.  This primtive picture allows one to derive a value for the mean-field quantum capacitance $C_q$, which in the mean-field approach imposes an upper limit on the observable $C(V)$. 

Before calculating the volumetric capacitance in the mean-field approach, one can consider first the problem of a single graphene sheet gated by a parallel metal electrode that is separated by a distance $d$ from the sheet (as in the experiments of Refs.\ \cite{Xia2009moq, Ponomarenko2010dos}).  Such a system can be described as a parallel-plate capacitor, and its total energy $\mathscr{U}$ per unit area is equal to $\mathscr{U}_q + \mathscr{U}_{el}$, where $\mathscr{U}_q$ is the quantum kinetic energy per unit area of electrons within the graphene and $\mathscr{U}_{el}$ is the electrostatic energy.

The quantum kinetic energy $\mathscr{U}_q$ can be obtained by integrating Eq.~\eqref{eq:nu_T}. This produces (at $T = 0$)
\begin{equation} 
\mathscr{U}_q = \frac23 \mu_e(n_e) n_e\,,
\label{eq:UAkMF}
\end{equation}
where
\begin{equation} 
\mu_e(n_e) = \hbar v \sqrt{\pi n_e}
\label{eq:mu_e}
\end{equation} 
is the chemical potential of graphene with the 2D electron concentration $n_e = \sigma/e$. In terms of the capacitor's charge per unit area $\sigma$, $\mathscr{U}_q$ becomes
\begin{equation} 
\mathscr{U}_q = \frac{\sqrt{e \sigma^3}}{6 \sqrt{\pi} \alpha \varepsilon_0 \varepsilon_r},
\label{eq:UqMF}
\end{equation}
where 
\begin{equation} 
\alpha = \frac{e^2}{4\pi \varepsilon_0 \varepsilon_r \hbar v}
\label{eq:alpha}
\end{equation}
is the dimensionless interaction constant of graphene.  For free-standing graphene $\alpha \approx 2.2$, so that for $\varepsilon_r = 3.0$ we have $\alpha \approx 0.7$.

Using the thermodynamic equations relating the energy $\mathscr{U}$ to the voltage $V$ and capacitance per unit area $\mathscr{C}$ produces an expression for the quantum capacitance per unit area $\mathscr{C}_q$.
In 2D the relations analogous to Eqs.~\eqref{eq:Vdef} and \eqref{eq:Cdef} are $V = d\mathscr{U}/d\sigma$ and $\mathscr{C} = (d^2\mathscr{U}/d\sigma^2)^{-1}$, so that the quantum capacitance is given by
\begin{equation} 
\mathscr{C}_q = 32 \pi \alpha^2 (\varepsilon_0 \varepsilon)^2 V/e.
\label{eq:CA-MF}
\end{equation}

This quantum capacitance is added in series with the geometric capacitance
\begin{equation}
\mathscr{C}_{g}(d)
 = \frac{\varepsilon_0 \varepsilon_r}{d}\,
\label{eq:Cg}
\end{equation} 
that results from the capacitor's electrostatic energy,
\begin{equation} 
\mathscr{U}_{el} = \frac{\sigma^2 d}
{2 \varepsilon_0 \varepsilon_r}\,.
\label{eq:UAelMF}
\end{equation}
Eqs.\ \eqref{eq:UqMF} and \eqref{eq:UAelMF} imply that when the charge of the capacitor is small enough that $\sigma \ll e/9\pi\alpha^2 d^2$ we have $\mathscr{U}_q \gg \mathscr{U}_{el}$, and therefore the quantum capacitance dominates the series sum $\mathscr{C} = (\mathscr{C}_q^{-1} +\mathscr{C}_g^{-1})^{-1} \simeq \mathscr{C}_q$.  Conversely, at $\sigma \gg e/9\pi\alpha^2d^2$ we have $\mathscr{C} \simeq \mathscr{C}_g$.

\begin{figure}[htb!]
\centering
\includegraphics[width=.5 \textwidth]{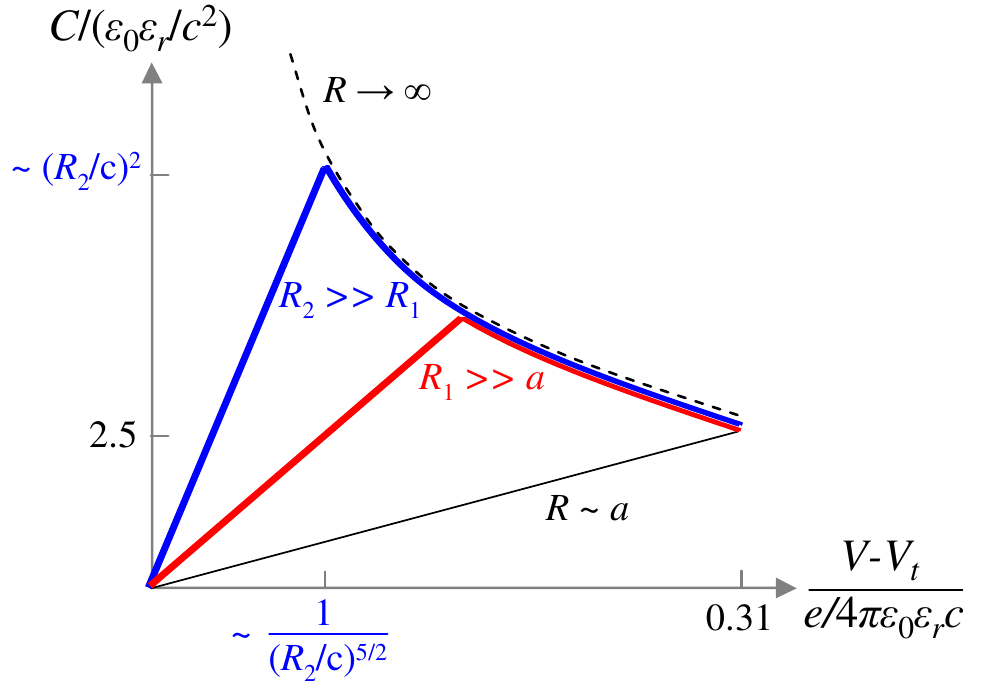}
\caption{The capacitance as a function of voltage for the hypothetical case where $\alpha = 1$ and the disk charge density $\sigma_0 = e/c^2$, plotted for different values of the disk size $R$.  At $R/a \sim 1$, the elastic energy-mediated attraction between ions is eliminated and the capacitance follows the mean-field quantum capacitance given by Eq.~\eqref{eq:CGS-MF} (thin, black line).  At $R \rightarrow \infty$, ions enter the GS in stages and the capacitance follows the result of Eq.~\eqref{eq:Csafran} (dashed curve).  The behavior for finite $R$, derived in Sec.~\ref{sec:alpha1}, is shown schematically by the red and blue thick lines for two different values of $R$.  Proper numerical coefficients are shown on the axes for the thin black and dashed curves, while for the thick curves the location and height of the peak capacitance is indicated only by an approximate scaling relation.} \label{fig:CV-GS-alpha1}
\end{figure}

For a stack of many graphene sheets with a uniform compensating ion charge, the mean-field picture suggests that the electrode consists of many such single graphene capacitors connected in parallel.  If one assumes that the distance between graphene sheets is a constant $c$, then the volumetric capacitance satisfies $C(V) = \mathscr{C}(V)/c$.  Substituting Eq.\ \eqref{eq:CA-MF} gives
\begin{equation} 
C_q(V) \simeq 8 \alpha^2  \frac{V - V_t}{e/4\pi\varepsilon_0 \varepsilon_r c} \frac{\varepsilon_0 \varepsilon_r}{c^2}.
\label{eq:CGS-MF}
\end{equation}
In principle, this quantum capacitance should be added in series with a constant geometric term $C_g$ that results from the system's electrostatic energy.  However, the quantum capacitance $C_q$ rises to the level of the geometrical value only at very large ion densities $N_+ \sim (\alpha^2c^3)^{-1}$, which are not physically realistic.  One can therefore say that the mean-field approach predicts a capacitance $C(V) \simeq C_q(V)$ over the entire relevant range of voltage.  $C_q(V)$ is plotted as the thin line in Fig.~\ref{fig:CV-GS-alpha1}.

We note here that Eq.\ \eqref{eq:Cflat} and all subsequent formulas for the volumetric capacitance are normalized to the volume of the empty GS, where graphene sheets are separated by a distance $c$.

%
%

\section{Capacitance of staged graphite} \label{sec:staging}

The mean-field picture presented in the previous section is apparently incompatible with the phenomenon of staging observed in graphite intercalation compounds, as described in Sec.~\ref{sec:GSelectrodes}. In staged graphite intercalated ions do not fill the electrode volume uniformly, but instead form a sequence of filled and empty interlayer galleries, periodic in the stacking direction $z$.
To compute the capacitance of this structure one can use the Thomas-Fermi (TF) approximation~\cite{Pietronero1978cdi, Safran1980eia}. The key findings of this approach are summarized below, with further details given in Appendix~\ref{sec:StagingDetails}.

In order to determine the energy of the staged arrangement of ions, one can approximate the ion layers as a sequence of uniform planes with surface charge density $\sigma_0$ positioned at
\begin{equation}
 z = (m + 1 / 2) h\,,
 \quad m = 0, \pm 1, \pm 2, \ldots,
\label{eq:z_planes}
\end{equation}
where $h$ is related to the average ion concentration $N_+$ by
\begin{equation}
h = \sigma_0 / (e N_+)\,.
\label{eq:h_from_N}
\end{equation}
In this description the discreteness of graphene sheets is also ignored, so that $h$ can take on values that are not integer multiples of $c$; this assumption is discussed at the end of this section.

Within the TF approximation, the electrostatic potential $\phi(z)$ and the 3D electron density $N_e(z) = n_e(z)/c$ are related by $e \phi(z) = \mu_e \bigl(n_e(z)\bigr) + \text{const}$. With the help of Eq.~\eqref{eq:mu_e} and a suitable choice of the additive constant, the Poisson equation for $\phi(z)$ can be written as
\begin{equation} 
\phi^{\prime\prime}(z) = \frac{16\pi\varepsilon_0\varepsilon_r \alpha^2}{e c} \phi^2(z)\,.
\label{eq:TF}
\end{equation}
By the Gauss law, $\phi^\prime(z)$ exhibits a discontinuity of $\sigma_0 / (2 \varepsilon_0\varepsilon_r)$ across each ionic plane.

For the case of large $h$, the solution for $\phi(z)$ can be approximated by~\cite{Pietronero1978cdi, Safran1980eia}
\begin{equation} 
\phi(z) \simeq \frac{3}{2\alpha^2}\, \frac{e}{4 \pi \varepsilon_0 \varepsilon_r}\,
          \frac{c}{(\Delta z + z_0)^2}\,,
\label{eq:phiTFplanes}
\end{equation}
where $\Delta z$ is the distance from the nearest ionic plane and
\begin{equation} 
z_0 = c \left(
 \frac{3}{2\pi\alpha^2} \frac{e}{\sigma_0 c^2}
  \right)^{1/3}
\label{eq:z0-GS}
\end{equation}
is the characteristic thickness of the plane's screening atmosphere.
The corresponding 3D electron density $N_e(z)$ near ionic planes is given by
\begin{equation} 
N_e(z) \simeq \frac{9 c}{4 \pi \alpha^2 (\Delta z + z_0)^4}.
\label{eq:n_near}
\end{equation}
For large ionic plane separation $h \gg z_0$ one can talk about an effective repulsion of the planes~\cite{Safran1980eia} due to the overlap of their screening atmospheres. The electron density perturbation caused by the overlap is the most significant near the midplanes, \textit{e.g.} at $z = 0$, where the density of the electrons and their screening ability are strongly diminished. Therefore, the interaction energy $u(h)$ per plane per unit area can be estimated as $u(h) \sim e \phi(0) N_e(0) h \propto 1 / h^5$. This estimate
is verified by a detailed calculation (cf.~Appendix~\ref{sec:StagingDetails}), which provides the numerical coefficient:
\begin{equation} 
u(h) \simeq \frac{c_1}{\alpha^4}
     \frac{e^2 c^2}{\varepsilon_0 \varepsilon_r h^5}\,,
     \quad
     c_1 = 1.16953\,.
\label{eq:ustage}
\end{equation}
The total energy per unit volume,
\begin{equation}
F \simeq \frac{u(h)}{h} + e V_t N_+\,,
\label{eq:F+_dilute}
\end{equation}
includes contributions from both the interplane repulsion and the self-energy of each plane, which is
\begin{equation}
u_0 = \left(
\frac{81 \pi^2}{250}\, \frac{\sigma_0^2}{e^2}\,
\alpha c \right)^{1/3}
 \hbar v
\label{eq:eV_t}
\end{equation}
per ion (cf.~Appendix~\ref{sec:StagingDetails}) and which enters into the threshold voltage $V_t$.
Using these results, we can derive the voltage $V$ and the capacitance $C$ of the staged compound near the intercalation threshold as a function of interplane distance $h$.  For large $h$ this is done by substituting the formula $N_+ = \sigma_0 / e h$ for the ion concentration and Eq.~\eqref{eq:F+_dilute} for the energy density
into the thermodynamic relations~\eqref{eq:Vdef} and \eqref{eq:Cdef}.
Next, eliminating $h$, we obtain
\begin{equation} 
C(V) \simeq 1.03 \, \frac{\varepsilon_0 \varepsilon_r}{c^2}
\left( \frac{\sigma_0 c^2}{e} \right)^{6/5}
\left(
\frac{\alpha e / 4\pi\varepsilon_0 \varepsilon_r c}{V - V_t} \right)^{4/5}.
\label{eq:Csafran}
\end{equation}

We should note that in the derivation of Eq.\ \eqref{eq:Csafran} it is assumed that the graphite can be represented as a continuous, electron-filled medium.  In a more realistic treatment, one could treat $h$ not as a continuous variable, but as a distance whose value is restricted to be an integer multiple of $c$, known as the stage number $s$ in graphite~\cite{Fischer1978gic, Dresselhaus1981ico}. In this case the ideal periodic structure discussed above would exist only for a set of discrete ion concentrations $N_+ = \sigma_0 / (e c s)$. At such concentrations $V$ would change discontinuously, giving a sharp peak in $C(V)$. In between the peaks, the system would be a mixture of two stages with neighboring stage numbers. Equation~\eqref{eq:Csafran} would then represent $C(V)$ averaged over an interval of voltage containing several peaks.

One could also notice that we have assumed ionic planes to be immersed in the continuum electron background without any gap between the two. If nonzero gaps of width $d_0 \sim a/2 + c$ exist at both sides of every plane, then the energy density acquires an extra term $e \sigma_0 N_+ d_0 / (\varepsilon_0 \varepsilon_r)$. Since $\sigma_0$ is a constant, this extra term can be absorbed into $V_t$:
\begin{equation}
V_t \to V_t + \frac{\sigma_0}{\mathscr{C}_{g}(2 d_0)}\,.
\label{eq:V_t_with_d}
\end{equation}
Notably, the ``geometric capacitance'' $\mathscr{C}_{g}$ [Eq.~\eqref{eq:UAelMF}] only changes the threshold voltage but does not reduce $C$, unlike in the mean-field theory of Eq.~\eqref{eq:C_EDL_MF} and Sec.~\ref{sec:CGS-MF}.

The behavior of $C(V)$ predicted by Eq.~\eqref{eq:Csafran} is markedly different from the mean-field theory Eq.~\eqref{eq:CGS-MF}, as shown by the dashed line in Fig.~\ref{fig:CV-GS-alpha1}. Most notably, the capacitance of staged graphite diverges at the threshold voltage rather than collapsing to zero, and it remains parametrically larger than the result given by Eq.~\eqref{eq:CGS-MF} up until $V - V_t \sim e / (4\pi\varepsilon_0 \varepsilon_r c)$. At those large voltages, the approximate formula of Eq.~\eqref{eq:Csafran} becomes inaccurate and
one needs a more careful method to calculate the capacitance, as outlined in Appendix~\ref{sec:StagingDetails}. The results of these calculations
are discussed in Sec.~\ref{sec:GSdiscussion}.

\section{Capacitance of a graphene stack} \label{sec:alpha1}

For the main problem of this paper, as defined in Sec.~\ref{sec:GSelectrodes}, the different results of Eq.~\eqref{eq:CGS-MF} and Eq.~\eqref{eq:Csafran} present something of a puzzle. One the one hand, when the spacers in the GS electrode are plentiful, the graphene sheets are fully separated and ions are free to enter the GS individually and fill its volume uniformly.  In this case, the capacitance should be similar to the mean-field result of Eq.~\eqref{eq:CGS-MF}. On the other hand, when the spacers in the GS are sparse, the intercalating ionic charges arrange themselves in the staged structure discussed in Sec.~\ref{sec:staging} and the capacitance should follow the prediction of Eq.~\eqref{eq:Csafran}.  In this section we show that both relations $C \propto (V - V_t)$ [as in Eq.~\eqref{eq:CGS-MF}] and $C \propto (V-V_t)^{-4/5}$ [as in Eq.~\eqref{eq:Csafran}] are realized in suitable ranges of voltage and we explain how the crossover between them occurs.

For the purpose of our conceptual discussion in this section we drop all numerical coefficients and focus instead on the scaling behavior of the capacitance as a function of voltage.  
For simplicity, we consider the case where $\alpha \sim 1$ and $a/c$ is also of order unity.  This second assumption amounts to assuming that $\sigma_0 \sim (e/a^2) \sim (e/c^2)$.  The effect of allowing $\alpha$ or $\sigma_0 / (e / c^2)$ to be small is discussed briefly at the end of the section, but we note here that as long as $a/c \ll (R/c)^{3/4}$ the scaling relations we derive below are unaffected.  A more thorough derivation of the capacitance is given in Ref.~\onlinecite{Skinner2011mto}, which includes an analysis of additional regimes that appear if $\alpha \ll 1$ or $\sigma_0 \ll e / c^2$.  

Our essential observation is that when disks of charge enter the GS, each disk draws around itself a strongly-bound, nonlinear screening atmosphere of electrons that renormalizes its charge.  To see how this happens, consider first the case where a single disk is introduced into the GS.
At very small distances $z$ from the surface of the disk, the disk behaves like an infinite plane and the electric potential can be described by Eq.~\eqref{eq:phiTFplanes}.  This formula for the potential is derived under the TF approximation, which neglects any quantum effects associated with finite electron wavelength, and therefore its applicability is limited to regions where the Fermi wavelength $\lambda_F$ is much smaller than the length scale over which the potential is varying in the transverse direction.  Thus, to find the distance $z_{\text{TF}}$ at which TF screening ends, one can equate $\lambda_F \sim 1/\sqrt{n_e} \sim e / (\varepsilon_0 \varepsilon_r \phi)$ to the disk size $R$, which by Eq.~\eqref{eq:phiTFplanes} gives 
\begin{equation} 
z_{\text{TF}} \sim \sqrt{R c}\,.
\end{equation}  
Since $z_{\text{TF}} \ll R$, the region over which the TF approximation is applicable can be described as a disk-shaped volume with thickness $z_{\text{TF}} \ll R$ and diameter $\sim R$.

The net effect of the strong, TF screening at $|z| < z_{\text{TF}}$ is to renormalize the charge of the disk. To see the extent of this renormalization, one can calculate the charge enclosed within the region $|z| < z_{\text{TF}}$.  In this case the Gauss law can be written as $q \sim -\varepsilon_0\varepsilon_r R^2 (d \phi/dz)|_{z = z_{\text{TF}}}$, which gives $q \sim e \sqrt{R / c}$.
This charge $q$ enclosed within the TF region is much smaller than the bare disk charge $\sigma_0 R^2 \sim e(R / c)^2$.

Outside of the TF region, the potential created by the reduced charge $q$ is not strongly screened.  Indeed, at these large distances the potential is affected only by the relatively weak dielectric response of graphene layers with a small TDOS.  Such linear screening is described in detail in Appendix~\ref{sec:Linear}. For the purpose of the present calculation of the capacitance, however, it is sufficient to think that outside of the TF region there is no screening of the disk charge.  Instead, the remaining negative charge $-q$ required to make the system electroneutral is distributed uniformly throughout the electrode. 

%
%
\begin{figure}[htb!]
\centering
\includegraphics[width=0.38 \textwidth]{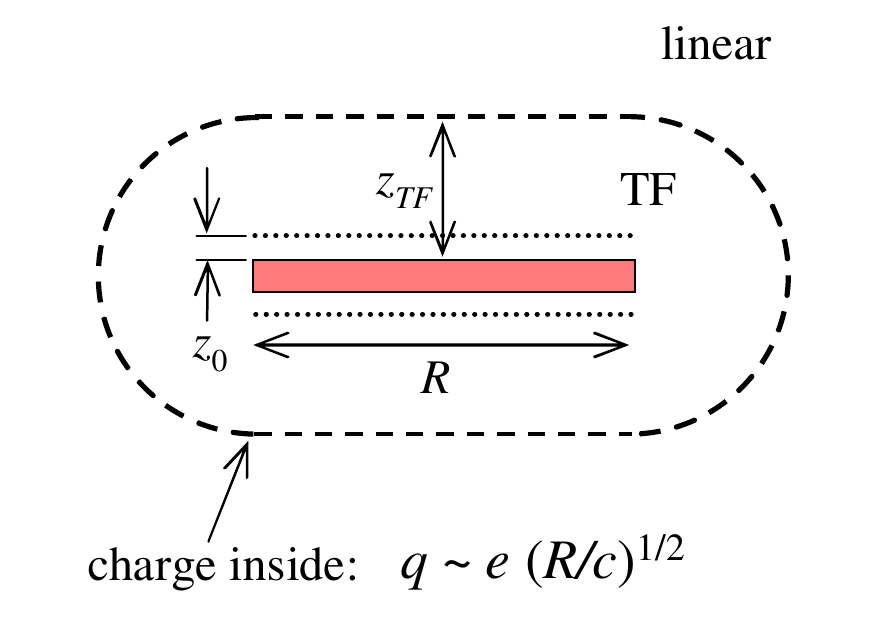}
\caption{A schematic depiction of the screening atmosphere surrounding a disk of ionic charge (rectangle) with diameter $R$ inside the graphene stack.  The limit of TF screening, which corresponds to a distance $z_{\text{TF}}$ from the disk, is shown by the dashed line.  The region outside the TF region corresponds to linear screening.  The characteristic decay length $z_0$ of the nonlinear potential is indicated by the dotted line.  The total amount of charge inside the TF screening atmosphere is denoted $q$ and is much smaller than the bare charge $\sigma_0 R^2 \sim e (R/c)^2$ of the disk of ions.} \label{fig:screening-alpha1}
\end{figure}

We can now consider what happens when there is some finite concentration $N$ of these charge-renormalized disks within the volume of the electrode.  When the concentration of the disks is low enough that $N \ll 1 / (R^2 z_{\text{TF}})$, the disks are well-separated from each other and their nonlinear screening atmospheres do not overlap.  One can think, then, that in this low density limit the capacitor charge consists of a dilute gas of disks, each with effective charge $q$, neutralized by a uniform background of electronic charge with charge density $-q N$.  

In this configuration, the total capacitor energy $U$ has three components: the self-energy of the disks with effective charge $q$, the electrostatic energy $U_{el}$ associated with the Coulomb interaction between the disks and the uniform background, and the quantum kinetic energy $U_q$ of the uniform, neutralizing electronic charge.  The self-energy component affects only the threshold voltage $V_t$ and does not enter into the capacitance.  The Coulomb energy $U_{el}$ is much smaller in magnitude than $U_q$ in the limit $N \ll 1 / (R^2 z_{\text{TF}})$; this is shown explicitly in Appendix~\ref{sec:Linear}.  Thus, in the regime where the disk screening atmospheres do not overlap, the capacitance is dominated by the quantum kinetic energy of the uniform electronic background.

The quantum kinetic energy $U_q$ can be derived in a way that is similar to the derivation of Eq.~\eqref{eq:UqMF}.  Here, the 3D electron density (in regions outside the TF screening atmospheres of the disks) is given by $N_e = q N / e$.   Using $q \sim e \sqrt{R / c}$ and Eq.~\eqref{eq:UAkMF} for the energy of graphene per unit area gives for the energy per unit volume $U_q \sim (N c^3)^{3/2} (R / c)^{3/4}(e^2/\varepsilon_0 \varepsilon_r c^4)$.  Since $N$ is related to the total capacitor charge $Q$ per unit volume by $Q = \sigma_0 R^2 N$, one can use the thermodynamic relations for capacitance and voltage in Eqs.~\eqref{eq:Vdef} and \eqref{eq:Cdef} to get
\begin{equation}
\begin{split}
C(V) \sim & \left(\frac{R}{c}\right)^{9/2}
                 \frac{V - V_t}{e/4\pi\varepsilon_0 \varepsilon_r c}                    \frac{\varepsilon_0 \varepsilon_r}{c^2}\,,
\\
& \textrm{if } \:
 \frac{V - V_t}{e/4\pi\varepsilon_0 \varepsilon_r c} \ll
 \left(\frac{c}{R}\right)^{5/2} .
\end{split}
\label{eq:Calpha1low}
\end{equation}
Corrections to this result associated with the Coulomb interaction between disks are discussed in Appendix~\ref{sec:Linear} [Eq.~\eqref{eq:CCoulombcorr}].

In Eq.~\eqref{eq:Calpha1low} one can see the strong effect of the renormalization of the disk charge.  This equation is similar the mean-field relation of Eq.~\eqref{eq:CGS-MF} in the sense that both give $C \propto (V-V_t)$, but Eq.~\eqref{eq:Calpha1low} has a significantly larger coefficient.   Indeed, the slope of the $C(V)$ relation in Eq.~\eqref{eq:Calpha1low} is larger than that of the mean-field result by a factor $(R/c)^{9/2} \gg 1$.  This large enhancement of the capacitance can be viewed as a direct result of the nonlinear screening of each disk.  The strongly-bound TF screening atmosphere surrounding each disk greatly reduces the uniform electron concentration in the electrode for a given capacitor charge $Q$ and therefore leads to a smaller quantum kinetic energy cost for capacitor charging.

When the capacitor charge is made larger, such that the concentration of disks $N \gg 1/(R^2 z_{\text{TF}})$, the nonlinear screening atmospheres of adjacent disks overlap and one can no longer talk about a uniform electron charge $-qN$ filling the space between disks.  Instead, the capacitance is dominated by the repulsive interaction between neighboring disks, and the capacitance is described by the staging theory of Eq.~\eqref{eq:Csafran}:
\begin{equation}
\begin{split}
C(V) \sim &  \left(
\frac{e/4\pi\varepsilon_0 \varepsilon_r c}{V - V_t} \right)^{4/5} \frac{\varepsilon_0 \varepsilon_r}{c^2},
\\
& \textrm{if} \: \frac{V - V_t}
                      {e/4\pi\varepsilon_0 \varepsilon_r c}
                       \gg
                  \left(\frac{c}{R}\right)^{5/2} .
\end{split}
\label{eq:Calpha1high}
\end{equation}
The resulting $C(V)$ dependence, which combines Eqs.~\eqref{eq:Calpha1low} and \eqref{eq:Calpha1high}, is shown schematically in Fig.~\ref{fig:CV-GS-alpha1}.  Notice that in the derivation of these results it was not necessary to assume ordering of the graphene sheets over distances larger than $R$.  Indeed, the low-density result of Eqs.~\eqref{eq:Calpha1low} assumes only that the concentration of graphene sheets is roughly uniform throughout the GS, while Eq.~\eqref{eq:Calpha1high} is based on disks interacting over a distance $z \ll z_{\text{TF}} \ll R$.  Therefore, our earlier assumption of long-range ordering of the GS does not significantly alter any results.

In this way the crossover between Eqs.~\eqref{eq:CGS-MF} and \eqref{eq:Csafran} can be understood as follows. At low voltages, one can still think that the result $C \propto (V - V_t)$ is a consequence of uniformly raising the electron Fermi level throughout the GS in order to provide a neutralizing electron concentration.  However, this neutralizing concentration should be thought of as a compensation not to the \emph{total} ionic charge but to the much smaller \emph{renormalized} ionic charge.  It is this renormalization that allows the capacitance to be large and produces a smooth crossover to the behavior $C \propto (V - V_t)^{-4/5}$ associated with staging.

The most dramatic consequence of this renormalization is the large peak of the capacitance at low voltage
$(V-V_t)/(e/4\pi\epsilon_0 \epsilon_r c) \sim (c/R)^{5/2}$, as shown in Fig.~\ref{fig:CV-GS-alpha1}.
Indeed at this point the capacitance attains a value which is larger than the maximum mean-field capacitance by the parametric factor $(R/c)^2$.  This factor can explain the large difference between the mean-field estimate and the large observed values of volumetric capacitance.  

In deriving the capacitance maximum we assumed that $\alpha$ and $\sigma_0/(e/c^2) \sim (c/a)^2$ were both of order unity.  If $\alpha$ or $\sigma_0$ are reduced, as should be expected when $a > c$, then naturally this peak in capacitance declines, since reducing either $\alpha$ or $\sigma_0$ implies a weaker role of Coulomb interactions relative to the quantum kinetic energy and therefore a weaker renormalization of the ion charge.  In fact, one can show \cite{Skinner2011mto} that as long as $\alpha \gg (c/R)$ and $\sigma_0/(e/c^2) \gg (c/R)^{3/2}$, the peak in capacitance of Fig.~\ref{fig:CV-GS-alpha1} should be replaced by the somewhat smaller value $\sim \alpha^2 (\sigma_0 c^2/e)^2 (R/c)^2 (\varepsilon_0\varepsilon_r/c^2)$, while the overall qualitative picture of $C(V)$ is not affected.

Finally, one can notice from Fig.~\ref{fig:CV-GS-alpha1} that a sparser arrangement of spacers within the GS, which corresponds to larger $R$, results in larger capacitance and therefore in greater energy storage for a given voltage.  However, such increased capacitance comes at the cost of slower capacitor charging.  That is, when spacers within the GS are sparse the process of ion intercalation into the GS is slow kinetically, as discussed in Sec.~\ref{sec:GSelectrodes}, and therefore the power of the device is reduced.  One can thus say that there is a fundamental tradeoff between high energy density and high power in graphene supercapacitors that can be adjusted by altering the density of spacers within the GS.

\section{Discussion} \label{sec:GSdiscussion}

In the previous section we showed how the capacitance can be much larger than the mean-field quantum capacitance $C_q$ as a result of the elastic energy-mediated attraction between ions and the nonlinear screening of disk-like ion bunches by the surrounding graphene layers.  These effects together produce a large, $R$-dependent peak in the capacitance at small voltages, as shown schematically in Fig.~\ref{fig:CV-GS-alpha1}.  

At larger voltages the capacitance is determined by the physics of staged graphite, as described in Sec.~\ref{sec:staging}.  
In the present section we discuss in greater detail the capacitance at these large voltages, and we show that near the steric limit of capacitor charging, where neighboring disks are separated by only a few graphene sheets, there is a noteworthy deviation of $C(V)$ from the result of Eq.~\eqref{eq:Csafran} toward larger capacitance.  This deviation brings the schematic picture of Fig.~\ref{fig:CV-GS-alpha1} closer in line with the large, mostly-constant $C(V)$ curve observed experimentally.

\begin{figure}
\centering
\includegraphics[width=0.45 \textwidth]{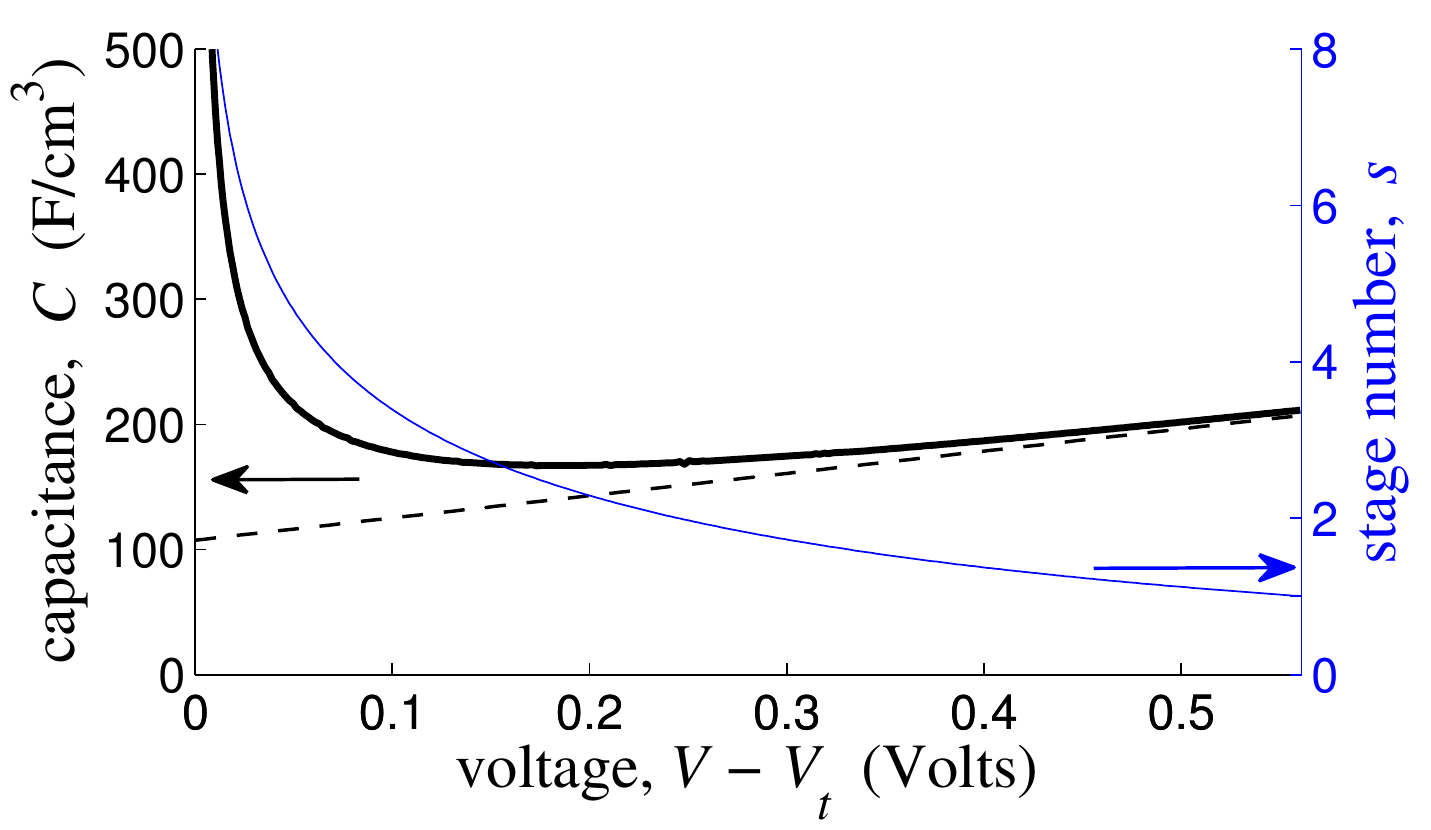}
\caption{(Color online) The volumetric capacitance $C$ (thick black curve, left axis) and the stage number $s$ (thin blue curve, right axis) as a function of voltage in the Thomas-Fermi approximation for turbostratic graphite ($R = \infty$), with $\sigma_0 = e/(1\,\text{nm})^{2}$, and $\varepsilon_r = 3$.  The capacitance in this plot is normalized to the volume of stage-1 graphite with intercalated ions of diameter $a = 1\,\text{nm}$.  The dashed black line is an approximation for $C(V)$ at large voltage, given by Eq.~\eqref{eq:Cflat}. } \label{fig:Safran}
\end{figure}

In our derivation of the approximate $C(V)$ relation of Eq.~\eqref{eq:Csafran}, it was assumed that disks of ionic charge are separated by a distance $h \gg z_0$, so that their TF screening atmospheres overlap only weakly.  This assumption becomes invalid when the capacitor charge is large enough that neighboring disks are separated by only a few graphene layers.  Instead, to calculate the capacitance at such large voltages one should employ the full solution of the TF equation [Eq.~\eqref{eq:TF}] rather than the asymptotic $1/h^5$ interaction law given in Eq.~\eqref{eq:ustage}.  Such a calculation is presented in Appendix~\ref{sec:StagingDetails}, and the results for capacitance and the stage number $s$ are shown in Fig.~\ref{fig:Safran} for $\varepsilon_r = 3$ and $\sigma_0 = e / (1\, \text{nm})^2$.  In this plot the capacitance is normalized to the volume of the filled electrode at stage $s = 1$ (which includes the volume of the ions as well as the GS itself), and assumes an ion size $a = 1\,\textrm{ nm} \approx 3 c$.  The right-most point of Fig.~\ref{fig:Safran} corresponds to stage 1, after which presumably no further capacitor charging is possible.

%
%

One can notice from Fig.~\ref{fig:Safran} that at small $V - V_t$ the capacitance declines with voltage according to the depedence $C \propto (V-V_t)^{-4/5}$ derived in Sec.~\ref{sec:staging}.  At larger voltages, however, the capacitance attains a minimum and then increases weakly with voltage.  This deviation from Eq.~\eqref{eq:Csafran} allows the capacitance to be larger and relatively constant over the majority of the operating voltage range.  

The origin of the upward deviation from Eq.~\eqref{eq:Csafran} can be understood as follows. At large enough voltages that $h < z_0$, the disks' screening atmospheres overlap strongly and the electron density becomes roughly uniform throughout the GS.  As a result, the energy at these large voltages is dominated by the quantum kinetic energy associated with uniformly raising the graphene Fermi level.  The capacitance therefore behaves similarly to the mean-field quantum capacitance described by Eq.~\eqref{eq:CGS-MF}, rising linearly with voltage.  The linear relation describing this portion of the $C(V)$ curve is shown as the dashed line in Fig.~\ref{fig:Safran}, and is derived below.  Its relatively small slope is identical to that of the mean-field quantum capacitance equation in Eq.~\eqref{eq:CGS-MF}, since at these large ion densities there is no renormalization of the ion charge.  Crucially, however, there is effectively a positive vertical offset of the quantum capacitance curve that allows the capacitance to be large. 

The value of this vertical offset can be calculated by first noting that the addition of a constant to Eq.~\eqref{eq:CGS-MF} is equivalent to substituting a shifted value of the threshold voltage $V_t$.  This shift in $V_t$ can be understood intuitively as the effective elimination of the self-energy of each disk.  That is, at $h < z_0$ each disk no longer has a well-defined screening atmosphere, so that there is no concept of a constant self-energy contributing to the threshold voltage.  Thus, an equation for the dashed line of Fig.~\ref{fig:Safran} can be obtained by replacing $V_t$ with $V_t - u_0/e$ in Eq.~\eqref{eq:CGS-MF}, where $u_0$ is the self-energy per ion of an isolated plane in the GS [Eq.~\eqref{eq:eV_t}]. Making this substitution gives the following expression:
\begin{eqnarray} 
C(V) & \simeq &
\left[ \left(\frac{20736\pi^2 \alpha^4}{125}\right)^{1/3} \right. \left(\frac{\sigma_0}{e/c^2}\right)^{2/3}  \nonumber \\
& & + \left. 8 \alpha^2  \frac{V - V_t}{e/4\pi\varepsilon_0 \varepsilon_r c} \right] \frac{\varepsilon_0 \varepsilon_r}{c^2}.
\label{eq:Cflat}
\end{eqnarray}
For situations where $z_0$ is as large as a few times $c$, such as may result when ions are large enough that $\sigma_0$ is relatively small, Eq.~\eqref{eq:Cflat} can occupy the majority of the SC's operating range of voltage.  This may help to explain why experiments on graphene SCs, where $z_0 / c$ is probably between $1$ and $3$, generally report a capacitance that varies only slightly with voltage and retains a value of $\sim 100$ F/cm$^3$ over the entire operating range of voltage \cite{Simon2008mec, Zhang2010gma, Zhang2010gnc, Kim2011hsb, Stoller2008gu, Wang2009sdb, Vivekchand2008ges, Zhu2011csp}.

At very small voltages, on the other hand, one should still expect a strongly-increasing, linear $C(V)$ relation associated with the quantum capacitance of the renormalized disk charge, as discussed in Sec.~\ref{sec:alpha1}.  This behavior leads to a large peak in the capacitance at a particular small voltage.
The apparent lack of such a sharp peak in experimental data is likely the result of finite temperature or disorder in the graphene stack.
Both disorder and finite thermal energy of ions work to diminish the positional correlations among disks of ionic charge when these disks are distant from each other, leading to a larger average interaction energy between disks and therefore to a smaller, somewhat smeared capacitance peak.  Nonetheless, even if positional correlations between disks are lost at stage $s > 4$ or so, as is commonly reported for traditional graphite intercalation compounds \cite{Dahn1991pdo}, Fig.~\ref{fig:Safran} suggests that a low-voltage capacitance peak of several hundred F/cm$^3$ may still be observable.

Long-range elastic interactions between ion bunches may also play an important role for the capacitance, and these will be explored in a later publication.

\begin{acknowledgments}

This work was supported primarily by the MRSEC Program of the National Science Foundation under Award Number DMR-0819885.  Additional support comes from NSF Grant No. PHY05-51164 and from UCOP.

\end{acknowledgments}

\appendix

\section{Thomas-Fermi theory of the staging regime} \label{sec:StagingDetails}

In this Appendix we provide the details of the TF theory of intercalated graphite discussed in Sec.~\ref{sec:staging}. It is convenient for our treatment here to introduce dimensionless variables $\bar\phi(z) = e \phi(z) / E_0$, $\bar{z} = z / \ell$, $\bar{z}_0 = z_0 / \ell$, and $\bar{h} = h / \ell$, where
\begin{equation}
E_0 = \hbar v \sqrt{\pi \sigma_0 / e}\,,
\quad
\ell = c \sqrt{\frac{3}{8 f^3}}\,,
\label{eq:E_0}
\end{equation}
are the energy and length units, and
\begin{equation}
f = \left(\frac{9\pi}{4}\, \frac{\sigma_0}{e}\,
          \alpha^2 c^2 \right)^{1/6} \sim 1
\label{eq:f}
\end{equation}
is a dimensionless parameter. In further calculations, it is sufficient to consider a single period $-\bar{h} / 2 < \bar{z} < \bar{h} / 2$,
so that Eq.~\eqref{eq:TF} yields the following boundary-value problem:
\begin{equation}
\bar\phi^{\prime\prime}(\bar{z}) = \bar\phi^2(\bar{z})\,,
\quad
\bar\phi^\prime(\pm \bar{h} / 2) = \pm \sqrt{2 f^3 /\, 3}\,.
\label{eq:phi_eq}
\end{equation}
Its solution can we represented by the inverse function
\begin{equation}
\bar{z}(\bar\phi) = \pm \frac{1}{\sqrt{\bar\phi_0}}\,
I\left(\frac{\bar\phi}{\bar\phi_0}\right)\,,
\quad
I(x) \equiv \int\limits_0^x
            \frac{d u}{\sqrt{\frac23 (u^3 - 1})}\,,
\label{eq:z_from_phi}
\end{equation}
where $\bar\phi_0$ stands for $\bar\phi(0)$. Note that
$I(x)$ can be expressed in terms of the Gauss hypergeometric function ${}_2F_1(a, b;\, c;\, x)$ and the Euler gamma function $\Gamma(x)$:
\begin{align}
I(x) &= I_\infty - {}_2F_1\left(
\frac16, \frac12;\, \frac76;\, \frac{1}{x^3}
\right) \sqrt{\frac{6}{x}}\,,
\label{eq:I}\\
I_\infty &= \sqrt{6\pi}\,\, \frac{\Gamma(7 / 6)}{\Gamma(2 / 3)}
= 2.97448\,.
\label{eq:I_infty}
\end{align}
For distant planes, $\bar h \gg 1$, we have $\bar\phi / \bar\phi_0 \gg 1$. Expanding $I(\bar\phi / \bar\phi_0)$ in Eq.~\eqref{eq:z_from_phi} to the leading order in this ratio and going through a simple algebra, we can show that near the ionic planes the potential behaves as
\begin{equation}
\bar\phi(\bar{z}) \simeq \frac{6}{\left(\frac12 \bar{h} - |\bar{z}| + \bar{z}_0\right)^2}\,,
\quad
\bar{z}_0 = \left(\frac{6}{f}\right)^{3 / 2}\,,
\label{eq:phi_near}
\end{equation}
which is equivalent to Eq.~\eqref{eq:phi_near}. The potential at the middle point $z = 0$ is given by
\begin{equation}
\bar\phi_0 \simeq 4 I_\infty^2 / \bar{h}^2\,,
\label{eq:phi_0_dilute}
\end{equation}
which implies that at this point Eq.~\eqref{eq:phi_near} errs by the factor $I_\infty^2 /\, 6 \approx 1.47$. This deviation is caused by the overlap of the screening atmospheres of the planes mentioned in Sec.~\ref{sec:staging}. The 3D electron density in physical units can be computed from
\begin{equation} 
N_e(z)  = \frac{\sigma_0}{e c}\, \bar\phi^2(\bar{z})\,,
\label{eq:n_near_II}
\end{equation}
which leads to Eq.~\eqref{eq:n_near}

Let us now compute the capacitance. First, we need to calculate the free energy density $F$ of the system. Keeping the electrostatic energy and electron kinetic energy but neglecting the entropy, we have
\begin{equation}
F = \frac{\sigma_0}{2 d}\, \phi\left(\frac{h}{2}\right)
+ \int\limits_{-h / 2}^{h / 2} \frac{d z}{h} e \phi(z) n(z) \left(
-\frac12 + \frac23\right).
\label{eq:F+_def}
\end{equation}
Evaluating the integral following Ref.~\onlinecite{Safran1980eia}, we get
\begin{equation}
F = \frac{\sigma_0 E_0}{5 e c}\, \left(
\frac{2}{\bar{h}}\, \bar\phi_h \sqrt{6 f^3} + \frac{\bar\phi_0^3}{3}
\right)\,,
\quad
\bar\phi_h = (f^3 + \bar\phi_0^3)^{1 / 3},
\label{eq:F+}
\end{equation}
where $\bar\phi_h$ denotes the potential at the ionic plane: $\bar\phi_h \equiv \bar\phi(\bar{h}/2)$.  Below we also use the shorthand notation $I_h \equiv I(\bar\phi_h / \bar\phi_0)$. Considering again the limit of large interplane distance, we can write the result of Eq.~\eqref{eq:F+} as
Eq.~\eqref{eq:F+_dilute} with
\begin{equation}
e V_t = \frac35\, E_0 f\,,
\label{eq:eV_t_II}
\end{equation}
which is equivalent to Eq.~\eqref{eq:eV_t}. The exact numerical coefficient in the interplane interaction energy $u(h)$ per unit area [Eq.~\eqref{eq:ustage}] is $c_1 = I_\infty^6 / (60 \pi^2)$, which is about $10\%$ smaller than what was obtained in Ref.~\onlinecite{Safran1980eia}.

In order to compute the voltage $V$
we take the derivative of $F$ with respect to the ion concentration:
\begin{equation}
eV = \frac{d F}{d N_+}
      = \frac{d F / d \bar\phi_0}{d N_+ / d \bar\phi_0}\,.
\label{eq:xi+_def}
\end{equation}
This concentration can be expressed in terms of our variables using
Eq.~\eqref{eq:z_from_phi}. The result is
\begin{equation}
N_+ = \frac{\sigma_0}{e h}\,,
\quad
s \equiv \frac{h}{c} = I_h \sqrt{\frac{3}{2 f^3 \bar\phi_0 \vphantom{\phi_0^\phi}}}\,,
\label{eq:N+}
\end{equation}
where $s$ is the dimensionless ``stage number.'' Equations~\eqref{eq:F+}, \eqref{eq:xi+_def}, and \eqref{eq:N+} imply
\begin{equation}
V = \frac{\hbar v \sqrt{\pi \sigma_0 / e}}{5} \left(3 \bar\phi
 + I_h \sqrt{\frac{6 \bar\phi_0^5}{f^3}}
  \right)\,.
\label{eq:xi+}
\end{equation}
Taking another derivative, we get the capacitance:
\begin{equation}
C_q = e\, \frac{d N_+}{d V}
= 2\pi \varepsilon_0 \varepsilon_r \alpha^2 \frac{\sigma_0}{e}\,
\frac{I_h \bar\phi_h^2 + \sqrt{6 f^3 \bar\phi_0 \vphantom{\phi_0^\phi}}}
     {I_h^3 \bar\phi_h^2 \bar\phi_0^2 \vphantom{\phi_0^\phi}}\,.
\label{eq:C_q_stage}
\end{equation}
It is easy to check that in the limiting cases of high and low ion concentrations $N_+$ we recover Eqs.~\eqref{eq:CA-MF} and \eqref{eq:Csafran}, respectively.

By virtue of Eqs.~\eqref{eq:I}, \eqref{eq:I_infty}, \eqref{eq:F+},
and \eqref{eq:N+}--\eqref{eq:C_q_stage}, all quantities of interest
are functions of $\bar\phi_0$. It is then possible to graph the dependencies of the capacitance $C_q$ and the stage number $s$ on voltage as parametric plots. An example is shown in Fig.~\ref{fig:Safran} and discussed in Sec.~\ref{sec:GSdiscussion}.

\section{Linear screening and renormalized mean-field theory in a graphene stack}
\label{sec:Linear}

In this Appendix we compute the screened potential of a Coulomb charge in the undoped GS and use it to calculate the correction to the renormalized mean-field theory expression for the volumetric capacitance [Eq.\ \eqref{eq:Calpha1low}].  We assume a relatively sparse filling of the GS by ions, so that the volume of the GS is not significantly expanded and we can still think of the GS as a stack of graphene sheets with separation $c$.

Due to the anisotropy of the system the screened electric potential $\phi_s(\rho, z)$ surrounding a point charge $q$
is anisotropic as well. Here $\rho$ is the radial coordinate in the $x$--$y$ plane. The starting point of the calculation is the electron dielectric function of the GS, which can be written as
\begin{equation} 
\varepsilon_e(k, k_z) = 1 - e\widetilde{\phi}_C(k, k_z) P(k, k_z)\,,
\end{equation}   
where $k = \sqrt{k_x^2 + k_y^2}\,$ is the in-plane momentum,
\begin{equation}
 \widetilde{\phi}_C(k, k_z) = \frac{q}{\varepsilon_0 \varepsilon_r}\,
 \frac{1}{k^2 + k_z^2}
\label{eqn:}
\end{equation}
is the Coulomb potential in the Fourier space, $P(k,k_z)$ is the polarization function, and henceforth the tilde marks the Fourier transform of the corresponding quanitity without the tilde.

If adjacent graphene layers are assumed to have no electronic interlayer coupling, as in turbostratic graphite, then $P(k, k_z)$ is related to the polarization function of 2D monolayer graphene, $P_2$, simply by $P(k, k_z) = P_2(k) / c$.  The polarization function $P_2(k)$ has been previously calculated~\cite{Gonzalez1994nfl, Ando2006sea} to be
\begin{equation} 
P_2 = - \frac14 \frac{|k|}{\hbar v} = - \frac{\pi \alpha \varepsilon_0 \varepsilon_r |k|}{e^2}.
\end{equation}
With this result one can define the potential $\widetilde{\phi}_s(k, k_z)$ in Fourier space:
\begin{align} 
\widetilde{\phi}_s(k, k_z) &= \frac{\widetilde{\phi}_C(k, k_z)}{\varepsilon_e(k, k_z)} 
= \frac{q}{\varepsilon_0 \varepsilon_r [k^2 +k_z^2 + \pi \alpha |k|/c]} \nonumber \\
 &\simeq \frac{q}{\varepsilon_0 \varepsilon_r [k_z^2 + \pi \alpha |k|/c]}\,.
\label{eq:philintilde}
\end{align}
The last, approximate relation in Eq.~\eqref{eq:philintilde} is valid for distances $\rho \gg c$, so that $k^2 \ll |k|/c$.  Taking the Fourier transform of this equation gives the following result for the potential:
\begin{equation} 
\phi_s(\rho, z) \simeq \frac{q}{4\pi \varepsilon_0 \varepsilon_r c} \left[ A_\rho^{-2/3} \frac{\rho}{c} + A_z^{-2/3} \frac{z^2}{c^2} \right]^{-3/2},
\label{eq:philin}
\end{equation}
where $A_\rho$ and $A_z$ are numeric coefficients of order unity, given by $A_\rho = \Gamma(3/4)^2/\sqrt{\pi^3 \alpha}$ and $A_z = 4/(\pi^2 \alpha^2)$. 

For the case of a disk of charge within the GS, Eq.~\eqref{eq:philin} can be used to describe the potential outside the TF screening atmosphere by substituting for $q$ the renormalized disk charge $q \sim e \sqrt{R/\alpha c}$.  It is worth noting that, to within numerical coefficients, Eq.~\eqref{eq:philin} smoothly matches the TF result of Eq.~\eqref{eq:phiTFplanes} at the boundary of the TF region.  That is, Eqs.~\eqref{eq:phiTFplanes} and \eqref{eq:philin} are equal at the points $\rho = 0$, $z = z_{\text{TF}} \sim \sqrt{R c/\alpha}$ and $z = 0$, $\rho \sim R$, which lie on the same equipotential contour.  Thus the potential surrounding a charged disk can be described by the TF result of Eq.~\eqref{eq:phiTFplanes} at $|z| < z_{\text{TF}}$, $\rho < R$ and by the linear response result of Eq.~\eqref{eq:philin} otherwise, with no parametric intermediate regime.

Using the linear potential of Eq.~\eqref{eq:philin}, one can calculate the Coulomb energy associated with a finite concentration of disks that are sufficiently separated from each other that their TF screening atmospheres do not overlap.
This Coulomb energy was ignored in our calculations of the capacitance in Sec.~\ref{sec:alpha1} at small $V - V_t$.  Indeed, the expression of Eq.~\eqref{eq:Calpha1low} is based on the quantum kinetic energy of electrons and neglects the electrostatic energy associated with the configuration of positively-charged disks residing on a negatively-charged background.  This approach was justified because when the concentration of disks $N$ is smaller than $1/(R^2 z_{\text{TF}})$ the Coulomb interaction energy between disks is much smaller than the quantum kinetic energy associated with the uniform electron charge.  In the remainder of this Appendix we explicitly calculate the Coulomb energy and prove this inequality.  We also find the small correction to the capacitance associated with the disks' Coulomb interaction; this is presented in Eq.~\eqref{eq:CCoulombcorr}.

As explained in Sec.~\ref{sec:alpha1}, when the concentration of disks is very small the capacitor charge consists of a sparse arrangement of charge-renormalized disks with charge $q$ and concentration $N$ surrounded by a uniform electron charge with density $-qN$.  The quantum kinetic energy per unit volume associated with the uniform electron charge is roughly
\begin{equation} 
U_q \sim (Nc^3)^{3/2} \left(\frac{R}{c}\right)^{3/4} \frac{e^2}{\varepsilon_0\varepsilon_r c^4},
\label{eq:Ukapp}
\end{equation}
as presented in Sec.~\ref{sec:alpha1}.
In the remainder of this Appendix, as in Sec.~\ref{sec:alpha1}, we drop all numeric coefficients and focus instead on parametric dependencies.  We also again assume that $\alpha \sim 1$ and $\sigma_0 \sim e/c^2$; the general case of small $\alpha$ and $\sigma_0$ is examined in Ref.~\onlinecite{Skinner2011mto}.

In order to estimate the magnitude of the Coulomb energy, one can consider that in their lowest energy configuration, the disks form a correlated arrangement on the uniform background such that the disks minimize their repulsive energy while maintaining the fixed concentration $N$.  This arrangement is characterized by the average spacing between disks in the $\rho$ and $z$ directions, which we denote $d_\rho$ and $d_z$, respectively.  The repulsive interaction between disks is dictated by the linear potential given in Eq.~\eqref{eq:philin}.  Since this potential is anisotropic, one can expect that the spacing between disks is also anisotropic.  In other words, the minimum energy arrangement of disks is that of an anisotropic Wigner crystal.  (This situation is similar to the better-studied system of colloidal particles that form a charge-renormalized 3D Wigner crystal \cite{Alexander1984cro}.)

The distances $d_\rho$ and $d_z$ can be found by noting that, in their minimum energy configuration, disks are arranged within the GS so that all nearest-neighbor interaction energies are equal in magnitude.  This implies that $d_\rho$ and $d_z$ are determined by the relation $\phi_s(d_\rho, 0) \sim \phi_s(0, d_z)$.  Since the concentration of disks $N \sim (d_\rho^2 d_z)^{-1}$, one can solve for $d_\rho$ and $d_z$ as a function of $N$.  This process gives $d_\rho/c \sim (N c^3)^{-2/5}$ and $d_z/c \sim (N c^3)^{-1/5}$, so that the typical nearest-neighbor interaction energy is $u_\textrm{nn} \sim q \phi_s(d_\rho, 0) =  q \phi_s(0, d_z) \sim (N c^3)^{3/5} (R/c)(e^2/\varepsilon_0\varepsilon c)$.

From the nearest-neighbor interaction energy $u_\textrm{nn}$ one can estimate the total Coulomb energy of the anisotropic Wigner crystal.  This Coulomb energy is, in fact, negative, as in the case of an ordinary isotropic Wigner crystal \cite{Mahan2000mpp}, since the attraction of each disk to its Wigner-Seitz cell of negative background charge is stronger than the repulsion between neighboring disks.  The magnitude of the Coulomb energy per disk is of the order of the nearest-neighbor interaction energy $u_\textrm{nn}$, and therefore the total Coulomb energy per unit volume $U_{el} \sim -N u_\textrm{nn}$, so that
\begin{equation} 
U_{el} \sim -(N c^3)^{8/5} \left(\frac{R}{c}\right) \frac{e^2}{\varepsilon_0\varepsilon c^4}.
\label{eq:Uelapp}
\end{equation}
Comparing Eqs.~\eqref{eq:Ukapp} and \eqref{eq:Uelapp} suggests that $|U_{el}| \ll |U_q|$ whenever $N \ll \sqrt{1/cR^5} \sim 1/(R^2 z_{\text{TF}})$.  Thus, our assumption that the Coulomb interaction is unimportant for the main term of the capacitance is justified.

Since the Coulomb energy of Eq.~\eqref{eq:Uelapp} is parametrically smaller than the quantum kinetic energy over the relevant range of voltage, $(V-V_t)/(e/4\pi\varepsilon_0\varepsilon_r c) \ll (c/R)^{5/2}$, its affect is only to provide a small correction to the main term of the capacitance, $C_q$, which is given by Eq.~\eqref{eq:Calpha1low}.
Specifically, since the Coulomb energy is negative and small, one can say that it produces a large negative capacitance per unit volume that is added in series with the relatively smaller main term.  This negative capacitance is an extension of the well-known negative compressibility of a conventional Wigner crystal \cite{Mahan2000mpp}, and it produces a small positive correction to Eq.~\eqref{eq:Calpha1low}.  Taking the appropriate derivatives of the total energy $U_q + U_{el}$ gives 
\begin{equation} 
C \sim C_q \left\{1 + \left[ \left(\frac{R}{c}\right)^{5/2} \frac{V-V_t}{e/4\pi\varepsilon_0\varepsilon_r c} \right]^{1/5} \right\},
\label{eq:CCoulombcorr}
\end{equation}
where $C_q$ is given by Eq.~\eqref{eq:Calpha1low}.
Notice that the correction term in Eq.~\eqref{eq:CCoulombcorr} grows to order unity precisely at the crossover point $(V-V_t)/(e/4\pi\varepsilon_0 \varepsilon c) \sim (c/R)^{5/2}$, where the TF screening atmospheres of neighboring disks begin to overlap and the capacitance transitions to the ``staging'' result $C \propto (V-V_t)^{-4/5}$ of Eq.~\eqref{eq:Csafran}.

It should be emphasized that the above derivation of Eq.\ \eqref{eq:CCoulombcorr} is schematic, and misses any numeric coefficients multiplying the Coulomb correction.  Such numeric coefficients may increase the magnitude of the Coulomb correction at small voltage and are potentially quite important.  A more careful calculation is the subject of a later publication.  Nonetheless, our main conclusion that $C$ should collapse to zero at $(V - V_t) = 0$, as does the quantum capacitance $C_q$, remains valid.

\bibliography{graphene_supercapacitors}
\end{document}